\documentstyle[epsfig,prl,aps,twocolumn]{revtex}
\newcommand{\be}{\begin{equation}}
\newcommand{\ee}{\end{equation}}
\newcommand{\bea}{\begin{eqnarray}}
\newcommand{\eea}{\end{eqnarray}}
\newcommand{\ba}{\begin{array}}
\newcommand{\ea}{\end{array}}

\newcommand{\FF}{{\cal F}_{P}}
\newcommand{\SS}{{\cal S}_{\vec{k} \vec{k}^{'}}}

\newcommand{\eps}{\epsilon}

\begin{document}
\draft
\title{\bf Random Scattering by Atomic Density Fluctuations in Optical 
Lattices}

\author{M. Blaauboer,$^{\rm a,b}$  
G. Kurizki, $^{\rm a}$  V.M. Akulin,$^{\rm c}$} 

\address{
$^a$ Chemical Physics Department, Weizmann Institute of Science, 
Rehovot 76100, Israel\\
$^b$ Department of Interdisciplinary Studies, Faculty of Engineering, 
Tel Aviv University, Tel Aviv 69978, Israel\\
$^c$ Laboratoire Aim\'{e} Cotton, CNRS II, B\^{a}timent 505, 
Orsay Cedex 91405 France}
\date{\today}
\maketitle

\begin{abstract} 
We investigate hitherto unexplored regimes of probe scattering 
by atoms trapped in optical lattices: weak scattering 
by effectively random atomic density distributions and 
multiple scattering by arbitrary atomic distributions. 
Both regimes are predicted to exhibit a universal 
semicircular scattering lineshape for large density fluctuations, 
which depend on temperature and quantum statistics. 
\end{abstract}


Recent advances in trapping and manipulation of cold atoms 
interacting with external fields have primarily been implemented 
thus far either in {\it single-atom} systems, such 
as sparse (low-density) optical lattices\cite{jess96}, or in Bose-Einstein 
condensates with {\it macroscopic} numbers of atoms\cite{park98}.
 
Between these two limiting regimes lies the scarcely investigated 
domain of
 processes involving a 
{\it finite number of interacting atoms}. 
Both the potential interest and the difficulties involved in studying
 
such processes are evident in the example of optical 
lattices with appreciable atomic 
filling factors\cite{depu99}, in
 
which the transition between "insulating" (localized) and "metallic" 
(superfluid) phases have been studied in the framework of the 
Bose-Hubbard model\cite{jaks97}. As confirmed by the above study, 
the mean-field approximation is inadequate for small numbers of 
interacting atoms in lattices due to the presence of large 
quantum fluctuations. Still more difficult is the analysis of fluctuations 
in systems of cold atoms coupled by long-range ($1/r$ or $1/r^3$) 
field-induced interactions\cite{odel99}. This leads to the intruiging 
question: Is there a way to circumvent the formidable 
task of treating the full dynamics of such systems and still infer 
their important characteristics, e.g., their dependence on temperature, 
quantum statistics (Bose or Fermi), number of atoms and lattice 
parameters? And: Are there universal measurable features 
which can be a "signature" of the statistical ensemble 
(distribution function) of such systems?

\begin{figure}
\centerline{\epsfig{figure=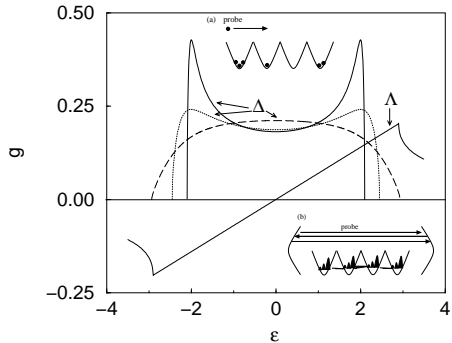,width=0.7\hsize}}
\caption[]{Density of states $g(\eps)$ of a probe scattered by bosonic atoms 
in a 1D optical lattice. Solid, dotted and dashed curves stand for 
$\Delta(\eps)$ and the thick curve stands for $\Lambda(\eps)$ 
(dispersion), see text. All curves 
are numerically computed from $G(\eps)$ and correspond to 
average random couplings $\langle W^2 \rangle$ = 0.4, 
2 and 10 respectively. The hopping frequency $J=1$, and for all 
curves $\int d\eps\, g(\eps) =1$. Inset a: A probe weakly scattered by a 
randomly occupied lattice. Inset b: A probe multiply scattered by a regular 
atomic distribution. 
} 
\label{fig:Dos1D} 
\end{figure}

Here we consider the possibility of inferring such statistical 
characteristics from the spectral features of probe photons or 
particles that are scattered by the density fluctuations of trapped atoms, 
notably in optical lattices, in two hitherto unexplored scenarios: (a) The 
probe is weakly (perturbatively) scattered 
by the local atomic density corresponding to the random occupancy 
of different lattice sites (Fig.~\ref{fig:Dos1D} - inset a). (b) The probe 
is multiply scattered by an arbitrary (possibly unknown a 
priori) multi-atom distribution in the lattice (Fig.~\ref{fig:Dos1D} - inset b). 

At the heart of our analysis is the idea 
that the Green function of the scattered photon or particle, which 
embodies the relevant spectral information, can be qualitatively 
estimated without resorting to cumbersome perturbative calculations 
of the probe-multiatom interaction by replacing this 
interaction Hamiltonian by an 
equivalent random matrix. The random matrix approach, which has been 
successfully applied to various disordered systems\cite{meht91}, 
allows the evaluation of probe spectra
 
to {\it all orders of scattering}, expressing them by means 
of {\it only the first two moments} (the mean and variance) 
of the random interaction, averaged over the statistical 
ensemble of the multiatom system. The highlight of our analysis, 
based on this random matrix approach, 
is the prediction of a semicircular spectral lineshape 
of the probe scattering in the large-fluctuation limit of 
trapped atomic ensembles. Thus far, the only known case of quasi-semicircular 
lineshapes in optical scattering has been predicted\cite{akul93} 
and experimentally verified\cite{ngo94} in dielectric microspheres with 
randomly distributed internal scatterers. 

The Green function of the probe (P) at energy $\eps = \hbar \omega$ 
is given by 
\be 
G_{P}(\eps) = \mbox{\rm Tr}_{S} \left[ \frac{1}{\eps-\hat{H}_{P} - 
\hat{V}}\, \hat{\rho}_{S} \right], 
\label{eq:Green}
\ee
where $\hat{H}_{P}$, $\hat{V}$ and $\hat{\rho}_{S}$ are, respectively,
 
the unperturbed probe Hamiltonian, the probe-system interaction  
Hamiltonian and the density operator for the ensemble of the multiatom system (S). 
We shall assume that the following conditions hold. (i)  
There is no appreciable back-effect of the probe on  
the multiatom system (otherwise it is no longer a probe). 
(ii) The state of the  
multiatom system does not change during the interaction time  
with the probe, i.e., the multiatom system remains "frozen", as
is applicable for optical or atomic probing. 
This situation then cannot be described as Markovian relaxation  
(exponential decay) of the probe state into the multiatom  
reservoir, since the 
correlation time of this reservoir is now much longer  
than that of the probe, in contrast with the basic assumption
 
of relaxation. (iii) The probe spectrum is broadband, i.e., it 
encompasses many of its eigenstates.

For an ensemble "frozen" during the interaction time, the tracing 
in (\ref{eq:Green}) implies statistical averaging over repeated 
realizations of the multiatom system, every time the probe scattering 
is recorded, or taking the expectation value with respect 
to the quantum state of the system. 
For simplicity, let us explicitly consider elastic scattering (the extension
 
to inelastic scattering is straightforward), for which
 
\begin{mathletters} 
\bea 
\hat{V} & = & \sum_{\vec{k}} f_{\vec{k}}\, \hat{\rho}^{\dagger}_{P \vec{k}}\, 
\hat{\rho}_{S \vec{k}} + \mbox{\rm h.c.} \hspace*{1cm} \mbox{\rm or} 
\label{eq:bilinear} \\  
\hat{V} & = & \sum_{\vec{k}} f_{\vec{k}}\, a^{\dagger}_{\vec{k}}\, 
\hat{\rho}_{S \vec{k}} + \mbox{\rm h.c.} 
\label{eq:linear} 
\eea 
\label{eq:potential} 
\end{mathletters} 
Here $f_{\vec{k}}$ is the scattering amplitude for momentum exchange  
$\hbar \vec{k}$ between the probe and the system and the $\vec{k}$-mode  
Fourier components of the probe (system) density operators  
$\hat{\rho}_{P \vec{k}}$ ($\hat{\rho}_{S \vec{k}}$) are defined 
in terms of their respective creation and annihilation operators 
$\hat{\rho}_{P \vec{k}} = \sum_{\vec{q}} a^{\dagger}_{\vec{q}} 
a_{\vec{q} + \vec{k}}$, $\hat{\rho}_{S \vec{k}} = \sum_{\vec{q}}  
c^{\dagger}_{\vec{q}} c_{\vec{q} + \vec{k}}$. Equations~(\ref{eq:bilinear})  
and (\ref{eq:linear}) 
stand, respectively, for bilinear and linear probe-system coupling. 
For optical probes (\ref{eq:bilinear}) and (\ref{eq:linear}) correspond 
to Raman and single-photon scattering, respectively. For atom or 
neutron probes the coupling (\ref{eq:bilinear}) is appropriate.

The Green function~(\ref{eq:Green}) is obtainable, to all orders in 
$\hat{V}$\cite{ande61}, by solving the set of equations for its diagonal elements 
\be 
G_{\vec{k} \vec{k}}(\eps) = [ \eps - \eps_{\vec{k}} - \sum_{\vec{k}^{'}} 
\langle \hat{V}_{\vec{k} \vec{k}^{'}}^2 \rangle\, G_{\vec{k}^{'} 
\vec{k}^{'}}(\eps) ]^{-1}, 
\label{eq:diagonal} 
\ee 
where $\eps_{\vec{k}}$ are the probe energy eigenvalues in the absence 
of potential fluctuations and pointed brackets denote the expectation value. 
The spectral information contained in these 
$G_{\vec{k} \vec{k}}$ is given by the density of states (DOS) of the probe 
$g(\eps) = - \frac{1}{\pi} \mbox{\rm Im} \sum_{\vec{k}} G_{\vec{k} 
\vec{k}}(\eps)$.

In order to extract information on the system we shall make two 
simplifying assumptions regarding the probe and the coupling 
potential~(\ref{eq:potential}): (i) $f_{\vec{k}}$ is flat in $\vec{k}$ (the 
coupling is strongly localized in space) within a band exceeding the 
relevant band of the system, so that $f_{\vec{k}} \approx f$; 
(ii) the statistical distribution of the probe is also flat in $\vec{k}$ and 
its second moment in $\langle \hat{V}^2 \rangle$ is replacable by 
the square of its mean flux (or density) $\bar{n}_{P}^2$ for the 
bilinear coupling (\ref{eq:bilinear}) or by its mean flux (density) 
$\bar{n}_{P}$ for the linear coupling 
(\ref{eq:linear}). Under these assumptions 
we can rewrite the squared coupling potential in (\ref{eq:diagonal}) as 
\be 
\langle \hat{V}_{\vec{k} \vec{k}^{'}}^2 \rangle = 
\langle \hat{V}_{\vec{k} \vec{k}^{'}} \rangle^2 +
\FF\, \SS.
\label{eq:coupl}
\ee 
Here $\langle \hat{V} \rangle$ is the mean coupling potential and $\FF \sim 
|f|^2\, \bar{n}_{P}^2$ or $\FF \sim |f|^2\, \bar{n}_{P}$ in the case of 
(\ref{eq:bilinear}) and (\ref{eq:linear}), respectively. The quantity 
of interest for the system is the Fourier-transformed 
density-density correlation of the atomic system 
\be 
\SS = \langle \hat{\rho}_{S\vec{k}}^{\dagger} \hat{\rho}_{S\vec{k}^{'}} 
\rangle + \mbox{\rm c.c}. 
\label{eq:S-matrix} 
\ee 
Its diagonal element ${\cal S}_{\vec{k} \vec{k}}$ is the static 
structure factor ${\cal S}_{\vec{k}}$, which is the Fourier transform  
of the van-Hove correlation function $\langle \hat{\rho}_{S}^\dagger(\vec{r},t=0) 
\hat{\rho}_{S}({\vec{r}}^{'},t^{'}=0) \rangle$ for the spatial density 
fluctuations of the "frozen" atomic ensemble.

The difficulty of having to evaluate or measure the matrix elements 
$\SS$ is avoided for a spatially random density distribution of the atomic 
system, due to random site occupancy (Fig.~\ref{fig:Dos1D}, inset a) and 
short-range interaction with the probe (e.g., a neutron or thermal atom). 
The elements $\SS$ in (\ref{eq:coupl}) and (\ref{eq:S-matrix}) can 
then be replaced by the average of the structure factor over all 
relevant $\vec{k}$: 
\be 
\SS \rightarrow \bar{\cal S} = \int d \vec{k}\, {\cal S}_{\vec{k}} \sim 
\langle \hat{n}_{S}^2 \rangle - \langle \hat{n}_{S} \rangle ^2, 
\label{eq:struct} 
\ee 
where the right-hand side of $\bar{\cal S}$ denotes the local atomic density 
or number variance averaged over the ensemble. 

The implications of evaluating the probe DOS $g(\epsilon)$ using 
(\ref{eq:diagonal})-(\ref{eq:struct}) will be examined for 
random fluctuations about a mean scattering potential $\langle V_{S}(x) \rangle$ 
(corresponding to the mean atomic density distribution) that is 1D-periodic. 
The "unperturbed" probe dispersion associated with $\langle V_{S}(x) \rangle$
is $\eps_{\vec{k}} = - 2J\, \cos (k_{x}d) + A$,  
$J$ being the hopping frequency, $d$ the lattice period and $A$ the band  
energy offset. This gives rise to the following expression for the 
Green function (\ref{eq:diagonal})
\be 
G(\eps) = \left[ \eps - \eps_{\vec{k}} - \langle W^2 \rangle
\sum_{\vec{k}^{'}} (\eps - \eps_{\vec{k}^{'}} - \Lambda (\eps)
+ i \Delta (\eps) )^{-1} \right]^{-1}.
\label{eq:Green2}
\ee
Here $\langle W^2 \rangle \equiv \FF \, \bar{{\cal S}}$, $\Lambda (\eps) =
\langle W^2 \rangle / \sqrt{(\eps - A)^2 - 4J^2}$ for 
$|\eps - A| > 2J$, $\Delta (\eps) = \langle W^2 \rangle /
\sqrt{4 J^2 - (\eps - A)^2}$ for $|\eps - A| < 2J$ and both zero
otherwise. Figure~\ref{fig:Dos1D} shows how the probe DOS $g(\epsilon)$ 
changes from that of a periodic band structure corresponding to the mean 
potential $\langle V_{S}(x) \rangle$ to a semicircular shape 
as the amount of fluctuations measured by 
$\langle W^2 \rangle$ increases. 

\begin{figure}
\centerline{\epsfig{figure=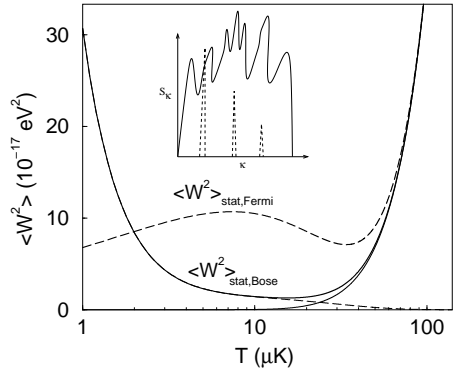,width=0.7\hsize}\vspace{0.5cm}}
\caption[]{Density-density fluctuations 
$\langle W^2 \rangle$ (in units of 10 (neV)$^2$) as a function of temperature T for 
bosonic and fermionic Li atoms in an optical lattice. Thin solid line - 
fluctuations due to evaporation (\ref{eq:evap}) (scaled by a factor of 7.8), 
thin dashed line - 
statistical fluctuations (\ref{eq:statBose}).  
Thick solid line - total fluctuations  
$\langle W^2 \rangle$ for bosonic Li atoms; thick dashed line - 
their counterpart for fermionic Li atoms. 
Parameters used are: $V_{0}=5$ neV, $\langle n_{S} \rangle$ = 0.1, d = 0.1 $\mu$m,  
$c_{p}$(Li) $\sim (3.6)\cdot 10^{6}$ J kg$^{-1}$  
K$^{-1}$, $\lambda_{F}$ (Li) = $6\cdot 10^{-10}$ m and  
$\omega_{v}$(Li)$\sim 2\cdot 10^6$ s$^{-1}$\cite{appr}. 
Inset: 
Static structure factor vs. $\kappa$ for phonons (solid curve) and  
nearly-free fermions (dashed curve) in a lattice at finite T.  
} 
\label{fig:bose} 
\end{figure}

In the multiple-scattering scenario,
 
which pertains to resonantly scattered atomic probes or to intracavity 
optical probes (Fig.~\ref{fig:Dos1D}, inset b), semicircular lineshapes are
 
obtained even when the $\SS$ cannot be claimed to belong to a random 
distribution (Fig.~\ref{fig:bose}, inset).
 
In the case of strongly-interacting atoms within a lattice site or longe-range 
intersite density correlations\cite{odel99} the distribution may be quite 
intricate, corresponding to sharp peaks of $\SS$. Nevertheless, the universal 
spectral trends of Fig.~\ref{fig:Dos1D} can be shown
 
to hold in this scenario, provided $\langle \hat{V}^2 
\rangle^{1/2} g_{0}(\eps) \gg 1$, $g_{0}(\eps)$ denoting 
the "unperturbed" probe DOS. This condition allows  
us to estimate $G_{\vec{k} \vec{k}}$ in (\ref{eq:diagonal}) to all orders in  
$\hat{V}$, upon replacing the state of the atomic system by a gaussian  
random ensemble\cite{meht91,akul93}. The result is the following universal 
formula\cite{akul93} for the renormalized probe energy $\tilde{\eps}$ at a 
given input energy $\eps$ 
\be 
\eps = \tilde{\eps} + \langle W^2 \rangle \mbox{\rm Tr}_{P} \left (
\frac{1}{\tilde{\eps} - \hat{H}_{P} - i0} \right).
\label{eq:renorm}
\ee
The use of (\ref{eq:renorm}) leads to a semicircular 
lineshape similar to the one in Fig.~\ref{fig:Dos1D}, as if the 
potential were random.

In order to illustrate the role of temperature, quantum statistics and 
the mean lattice potential in producing the semicircular lineshape, we 
proceed to evaluate $\langle W^2 \rangle$ for several simple models:
\newline  
1. \underline{The isolated-site limit}: The tightly-bound Bose or 
Fermi distributions in a lattice can be estimated by taking the 
potential of each site to be that of a harmonic well of depth $V_{0}$. 
The isolated-site approximation holds for atoms in the lowest vibrational 
band, when the coupling energy is much smaller than the 
excitation energy to the next band\cite{jaks97}, $\sqrt{\langle W^2 \rangle} 
\ll \hbar \omega_{\nu} = \frac{2\pi \hbar}{\lambda} \ 
\sqrt{\frac{2 V_{0}}{m}}$, $\lambda$ being the wavelength of the laser light. 

In the absence of additional 
external perturbations, the coupling $\langle W^2 \rangle$ arises  
because of temperature-dependent fluctuations in the site-occupancy of 
the optical lattice, which has an 
approximately gaussian distribution\cite{depu99}. 
The resulting random coupling energies averaged over all states 
yield $\langle W^2 \rangle  \sim  \FF \left\langle \langle n_{\vec{k}}, 
n_{\vec{k}^{'}} \rangle - \langle n_{\vec{k}} \rangle 
\langle n_{\vec{k}^{'}} \rangle \right\rangle_{\vec{k}\vec{k}^{'}}  
= \FF \left\langle \langle n_{i}, 
n_{j} \rangle - \langle n_{i} \rangle 
\langle n_{j} \rangle \right\rangle_{ij}  
\approx \FF \left( \langle n_{S}^2 \rangle - \langle n_{S}\rangle^2 
\right)$. Here $n_{\vec{k}} \equiv c^{\dagger}_{\vec{k}} c_{\vec{k}}$, $i$ and $j$ 
label atomic sites, and $\langle n_{S} \rangle$ is the average  
number of atoms per site. The last step 
applies whenever $n_{i} \approx n_{j}$ $\forall i,j$ 
and the density fluctuations are approximately site-independent. 
We have verified this by numerical simulation, considering 
2 to 4 identical atoms on a 1D lattice with  
6 sites and calculating the density fluctuations if the  
probability of an occupied site is 1/10 of the  
probability of an empty site. In all cases the maximum relative difference between 
$\FF \SS$ and $\langle W^2 \rangle$ was less than 10 \%.

The kinetic contribution to $\langle W^2 \rangle$ due to evaporation 
of atoms from the lattice is the dominant one at high temperatures, 
regardless of the statistics. If all the atoms are in the lowest energy  
band, we may adopt the rate equation used to describe the formation of 
electron-hole clusters in a plasma\cite{klin81} and find 
\be 
\langle W^2 \rangle_{\rm evap} = a\, \langle n_{S} \rangle\, T^2\, 
e^{-\beta V_{0}}. 
\label{eq:evap}
\ee
Here $a = k_{B} m c_{p}$, with $k_{B}$ the Boltzmann constant, $m$ 
the mass of the atoms and $c_{p}$ their specific heat, 
$T$ denotes the temperature, $\beta^{-1} \equiv k_{B} T$ and 
$V_{0}$ is the optical lattice potential. 
The influence of  
evaporation becomes the dominant effect for $T\sim 25\, \mu$K. 
Around $T\sim 300\, \mu$K 
these fluctuations become comparable in size to the square of the optical 
lattice potential ($\langle W^2 \rangle \sim V_{0}^2 \sim 100$ (neV)$^2$) 
and atoms then largely escape from the lattice. 

At low temperatures (well below 100 $\mu$K) the density-density fluctuations 
depend on whether the atoms in the lattice are bosons or fermions. 
For bosonic atoms in the lowest vibrational state we obtain\cite{reic98}
\be
\langle W^2 \rangle_{\rm stat,Bose} = \frac{z}{1-z} +
\left( \frac{z}{1-z} \right)^2 + \frac{d^3}{\lambda_{T}^3} 
\sum_{\alpha = 1}^{\infty} \frac{z^{\alpha}}{\alpha^{1/2}}.
\label{eq:statBose}
\ee
Here we have approximated the motion of the atoms in the potential wells 
by a harmonic oscillation with frequency $\omega_{v}$\cite{appr}, 
$z\equiv e^{- k_{B} T/\hbar \omega_{v}}$, $d$ denotes the average 
lattice spacing and $\lambda_{T} = (2\pi\hbar^2/m k_{B} T)^{1/2}$, the thermal 
wavelength, is the length scale separating quantum 
statistical behavior (for $\lambda_{T} \sim d$) from classical 
Maxwell-Boltzmann behavior (for $\lambda_{T} \ll d$). 
For fermionic atoms in an optical lattice\cite{andr99} one starts with 
the analog of the coupling~(\ref{eq:potential}) for particles
 
with spin, using creation and annihilation operators 
$c_{\vec{k}\sigma}^{\dagger}$ and $c_{\vec{k}\sigma}$ and 
performing an additional sum over the spin index $\sigma$, and 
follows the same analysis as above. One then finds  
\be
\langle W^2 \rangle_{\rm stat,Fermi} = \frac{z}{1+z} +
\left( \frac{z}{1+z} \right)^2 + \frac{2\, d^3}{\lambda_{T}^2 \lambda_{F}}\, 
\sum_{\alpha = 1}^{\infty} \frac{z^{\alpha}}{\alpha^{1/2}},
\label{eq:statFermi} 
\ee
with $\lambda_{F}$ the Fermi wavelength. At high temperatures $z \rightarrow 0$ 
and both (\ref{eq:statBose}) and (\ref{eq:statFermi}) reduce to the classical  
Maxwell-Boltzmann result $\langle W^2 \rangle_{\rm stat,clas} = 
z$.  
At low temperatures, 
fermionic fluctuations approach a constant value,  
whereas bosonic fluctuations become very large as $T$ decreases  
below $\sim 1 \mu$K, marking the Bose-Einstein condensation. 

In Fig.~\ref{fig:bose} we have taken typical 
parameters for available optical lattices\cite{appr,birk95,depu99} 
to show how $\langle W^2 \rangle$  
evolves as a function of temperature for both bosonic and 
fermionic Li atoms. The total density-density fluctuations consist 
of the sum of (\ref{eq:evap}) and either (\ref{eq:statBose}) or 
(\ref{eq:statFermi}), depending on the statistics. 
The isolated-site condition 
is satisfed for the entire temperature range displayed in 
Fig.~\ref{fig:bose}. Since the hopping frequency $J\sim V_{0}$, 
the random coupling for bosonic Li atoms changes from $\langle W^2 
\rangle/J^2 \sim\, 
10$ to  $\langle W^2 \rangle/J^2 \sim\, 100$, when going from $T\sim 8\,  
\mu$K to $T\sim 100\, \mu$K. Simultaneously the DOS 
then evolves from the periodic to the semicircular shape as in 
Fig.~\ref{fig:Dos1D}. The behavior for Na or Rb atoms 
is found to be similar to that of bosonic Li atoms , apart from their
different mass values in (\ref{eq:evap}) and (\ref{eq:statBose}). 
Employing currently achievable Bragg scattering techniques\cite{birk95} 
with a far off-resonant laser (1 mW/cm$^2$ intensity, 5.2 MHz detuning) 
and a lattice with a sufficiently high atomic filling factor\cite{depu99}, 
the scattering spectrum is expected to evolve with T, in the microkelvin 
range, from the  discrete to the semicircular regime.
2. \underline{The nearly-free limit}: A Bose or Fermi gas weakly modulated  
by the lattice potential yields 
$\SS = {\cal S}_{\vec{\kappa} = \vec{k} - \vec{k}^{'}} =  
| \phi_{\vec{\kappa}} |^2 {\cal S}_{\vec{\kappa}}^{(\mbox{\rm free})}$. 
Here $\vec{\kappa}$ is a reciprocal lattice vector, $\phi_{\vec{\kappa}}$ 
is the corresponding Fourier harmonic of the lattice potential (normalized to 1)  
and ${\cal S}_{\vec{\kappa}}^{(\mbox{\rm free})}$ is the structure factor  
for momentum transfer $\hbar \kappa$ in a free Bose or Fermi gas. For a  
Fermi gas ${\cal S}_{\vec{\kappa}}^{(\mbox{\rm free})} = \pm \Theta  
(k_{f} - \kappa)$, the Fourier transform of pair correlations with parallel 
or anti-parallel spins (which determines the sign): it is the well-known  
step function which vanishes for $\kappa$ larger than the Fermi wavevector $k_{f}$. 
At finite temperatures this distribution broadens. The 
replacement of the nearly-free fermionic $\SS$ by the average value 
(\ref{eq:struct}) is then justifiable only in the 
multiple-scattering scenario, while in the weak-scattering scenario the 
lattice potential harmonics $\phi_{\vec{k}}$ pick out well-defined  
${\cal S}_{\vec{k} - \vec{k}^{'} = \vec{\kappa}}$  
(Fig.~\ref{fig:bose}, inset - dashed line). 
\newline 
3. \underline{The phonon regime}: Excitations at frequencies  
below the chemical potential of a Bose condensate trapped in a lattice 
can produce collective phonon modes\cite{stam99} whose "frozen" spectrum is  
characterized by ${\cal S}_{\vec{\kappa}} = \sum_{\vec{q}} [ ( \langle  
n_{\vec{q}} \rangle + 1) \sum_{\vec{G}} \delta (\vec{\kappa} - \vec{q} - \vec{G}) 
+  \langle n_{\vec{q}} \rangle \sum_{\vec{G}} \delta (\vec{\kappa} + \vec{q} + 
\vec{G}) ]$, where $\langle n_{\vec{q}} \rangle$ is the mean number of phonons 
at temperature $T$ with wavevector $\vec{q}$, and $\vec{G}$ denotes the  
reciprocal lattice vector. The phonon mode spectrum includes quasi-local modes 
in the case of fluctuating atomic distributions. This naturally leads to the  
limit (\ref{eq:struct}) and an effectively random coupling  
(Fig.~\ref{fig:bose}, inset - solid line). 

To conclude, we have identified novel regimes of probe scattering by  
atoms trapped in optical lattices in the random-density and multiple-scattering  
regimes. These regimes cannot be treated by the mean-field approximation, but  
are characterized by a universal feature of large density fluctuations,  
namely, semicircular scattering lineshapes. This is the atom-optical  
analog of the semicircular broadening of the DOS in disordered electronic 
systems, which as far as we know has not yet been 
observed unambiguously. The observation of this atom-optical counterpart 
presents a nontrivial but feasible challenge for experimentalists.

This work was supported by US-Israel BSF , the Israel Council 
for Higher Education, Minerva and Arc-en-Ciel.

\end{document}